\documentclass[11pt]{article}

\usepackage{amsmath}
\usepackage{amsfonts}
\usepackage{amssymb}
\usepackage{times}
\usepackage{latexsym}
\usepackage{epsfig}
\usepackage{amsthm}
\usepackage{natbib}
\usepackage{hyperref}
\usepackage{color}

\renewcommand{\baselinestretch}{1.15}
\begin{document}

\title{An Active Set Algorithm to Estimate Parameters in \\ Generalized Linear Models with Ordered Predictors}

\author{Kaspar Rufibach$^{(1)}$ \\
$^1$
Universit\"at Z\"urich \\
Institut f\"ur Sozial- und Pr\"aventivmedizin \\
Abteilung Biostatistik \\
Hirschengraben 84 \\
8001 Z\"urich \\*[0.2cm]}

\maketitle

\begin{abstract}
In biomedical studies, researchers are often interested in assessing the association between one or
more ordinal explanatory variables and an outcome variable, at the same time adjusting for covariates of any
type. The outcome variable may be continuous, binary, or represent censored survival times. In the absence of
precise knowledge of the response function, using monotonicity constraints on the ordinal variables improves
efficiency in estimating parameters, especially when sample sizes are small. An active set algorithm that can
efficiently compute such estimators is proposed, and a characterization of the solution is provided. Having an efficient
algorithm at hand is especially relevant when applying likelihood ratio tests in restricted generalized linear
models, where one needs the value of the likelihood at the restricted maximizer. The algorithm is illustrated on
a real life data set from oncology.
\end{abstract}


\let\proglang=\textsf
\newcommand{\pkg}[1]{{\fontseries{b}\selectfont #1}}
\let\code=\texttt

\newcommand{\LLj}{\LL_j}
\newcommand{\JJcp}{\JJ_{c,p}}

\newcommand{\mycite}[1]{{\small \sc \citeNP{#1}}}

\setlength{\marginparsep}{0mm}
\newcommand{\kr}[1]{{\noindent \bf \color{red}#1}}       
\newcommand{\red}[1]{{\noindent \color{red}#1}}
\newcommand{\blanco}[1]{ }

\def\bs{\boldsymbol}
\newcommand{\ED}{{\mathbb F}_n}                     
\newcommand{\Xin}{X_1,\ldots,X_n}                   
\newcommand{\mean}{\frac{1}{n} \sum_{i=1}^n}        %
\newcommand{\sumn}{\sum_{i=1}^n}                    %
\def\d{\, \mathrm d}                                
\newcommand{\norm}[1]{\left\Vert #1 \right\Vert}    
\newcommand{\SupN}[2]{\Vert #1 \Vert_\infty^{#2}}   
\def\Var{\mathop{\rm Var}\nolimits}
\def\Bias{\mathop{\rm Bias}\nolimits}
\def\Mse{\mathop{\rm Mse}\nolimits}
\def\roc{\mathop{\rm ROC}\nolimits}
\def\auc{\mathop{\rm AUC}\nolimits}

\def\est{{\hat f}_n}                                
\def\lest{{\hat \varphi}_n}                         
\def\dest{{\hat F}_n}                               
\def\hest{{\hat \lambda}_n}                         
\newcommand{\Hol}[3]{\HH^{#1,#2}(#3)}               
\newcommand{\knots}[1]{\mathop{\rm knots(#1)}\nolimits}

\def\toD{\to_{d}}
\def\eqD{\stackrel{d}{=}}
\def\toP{\to_{\rm p}}
\def\toas{\stackrel{\mathrm{as}}{\to}}

\newcommand{\ve}[1]{\bf{#1}}
\newcommand{\mat}[1]{\ve{#1}}       

\def\iid{i.i.d.~}
\newcommand{\ie}{i.e.~}
\newcommand{\eg}{e.g.~}
\newcommand{\wrt}{w.r.t.~}
\newcommand{\as}{a.s.~}

\newcommand{\ruck}[1]{\strut\hspace{#1cm}}

\def\A{{\bf A}}
\def\B{{\bf B}}
\def\C{{\bf C}}
\def\D{{\bf D}}
\def\E{{\bf E}}
\def\F{{\bf F}}
\def\G{{\bf G}}
\def\H{{\bf H}}
\def\J{{\bf J}}
\def\K{{\bf K}}
\def\M{{\bf M}}
\def\O{{\bf O}}
\def\P{{\bf P}}
\def\S{{\bf S}}
\def\T{{\bf T}}
\def\U{{\bf U}}
\def\V{{\bf V}}
\def\W{{\bf W}}
\def\x{{\bf x}}
\def\X{{\bf X}}
\def\y{{\bf y}}
\def\Y{{\bf Y}}
\def\z{{\bf z}}

\def\AA{{\cal A}}
\def\BB{{\cal B}}
\def\CC{{\cal C}}
\def\DD{{\cal D}}
\def\EE{{\cal E}}
\def\FF{{\cal F}}
\def\GG{{\cal G}}
\def\HH{{\cal H}}
\def\II{{\cal I}}
\def\JJ{{\cal J}}
\def\KK{{\cal K}}
\def\LL{{\cal L}}
\def\MM{{\cal M}}
\def\NN{{\cal N}}
\def\OO{{\cal O}}
\def\PP{{\cal P}}
\def\QQ{{\cal Q}}
\def\RR{{\cal R}}
\def\SS{{\cal S}}
\def\TT{{\cal T}}
\def\UU{{\cal U}}
\def\VV{{\cal V}}
\def\WW{{\cal W}}
\def\XX{{\cal X}}
\def\YY{{\cal Y}}
\def\ZZ{{\cal Z}}

\def\alp{\alpha}
\def\gam{\gamma}
\def\Gam{\Gamma}
\def\del{\delta}
\def\Del{\Delta}
\def\eps{\varepsilon}
\def\kap{\kappa}
\def\lam{\lambda}
\def\Lam{\Lambda}
\def\sig{\sigma}
\def\Sig{\Sigma}
\def\th{\theta}
\def\Th{\Theta}
\def\om{\omega}
\def\Om{\Omega}

\def\bea{\begin{eqnarray*}}
\def\eea{\end{eqnarray*}}
\def\be{\begin{equation}}
\def\ee{\end{equation}}
\def\bean{\begin{eqnarray}}
\def\eean{\end{eqnarray}}
\def\barr{\begin{array}}
\def\earr{\end{array}}
\def\bdes{\begin{description}}
\def\edes{\end{description}}
\def\bi{\begin{itemize}}
\def\ei{\end{itemize}}

\def\nin{\noindent}
\def\nn{\nonumber}
\def\npb{\nopagebreak}

\def\comp{^{\rm c}}
\def\trans{^\top}
\def\cd{ \, \vert \, }
\def\Bcd{ \, \Big\vert \, }

\def\Bl{\Bigl}
\def\Br{\Bigr}
\def\la{\leftarrow}
\def\op{o_{\rm p}}
\def\Op{O_{\rm p}}
\def\da{\downarrow}
\def\ua{\uparrow}
\def\by{\times}
\def\til{\widetilde}
\def\hat{\widehat}
\def\setm{\setminus}
\def\subs{\subset}
\def\sups{\supset}
\def\lg{\langle}
\def\rg{\rangle}

\def\Ex{\mathop{\rm I\!E}\nolimits}
\def\Pr{\mathop{\rm I\!P}\nolimits}

\def\N{\mathbb{N}}
\def\Q{\mathbb{Q}}
\def\R{\mathbb{R}}
\def\Z{\mathbb{Z}}

\newcommand{\dom}{\mathrm{dom}}
\def\const{\mathop{\rm const}}
\def\arctanh{\mathop{\rm arctanh}}
\def\artanh{\mathop{\rm artanh}}
\def\argmax{\mathop{\rm arg\,max}}
\def\argmin{\mathop{\rm arg\,min}}
\def\ave{\mathop{\rm ave}}
\def\Beta{{\rm Beta}}
\def\Bin{{\rm Bin}}
\def\Borel{{\rm Borel}}
\def\cl{{\rm cl}}
\def\conv{\mathop{\rm conv}\nolimits}
\def\cone{\mathop{\rm cone}\nolimits}
\def\Cov{\mathop{\rm Cov}\nolimits}
\def\Corr{\mathop{\rm Corr}\nolimits}
\def\det{\mathop{\rm det}}
\def\diag{\mathop{\rm diag}\nolimits}
\def\diam{\mathop{\rm diam}\nolimits}
\def\dist{\mathop{\rm dist}\nolimits}
\def\dim{\mathop{\rm dim}\nolimits}
\def\Exp{\mathop{\rm Exp}\nolimits}
\def\extr{\mathop{\rm extr}\nolimits}
\def\Geom{\mathop{\rm Geom}\nolimits}
\def\Hyp{{\rm Hyp}}
\def\id{\mathop{\rm id}\nolimits}
\def\Leb{\mathop{\rm Leb}\nolimits}
\def\lin{{\rm lin}}
\def\Med{\mathop{\rm Med}\nolimits}
\def\Median{{\rm Median}}
\def\Poiss{{\rm Poiss}}
\def\rang{{\rm rang}}
\def\rank{{\rm rank}}
\def\sign{\mathop{\rm sign}}
\def\spann{{\rm span}}
\def\Std{\mathop{\rm Std}\nolimits}
\def\supp{{\rm supp}}
\def\trace{\mathop{\rm trace}}
\def\TV{\mathop{\rm TV}\nolimits}
\def\Var{\mathop{\rm Var}\nolimits}
\def\Mean{\mathop{\rm Mean}\nolimits}

\def\et{\quad\mbox{and}\quad}
\def\und{\quad\mbox{und}\quad}
\def\oder{\quad\mbox{oder}\quad}
\def\fuer{\quad\mbox{f"ur }}
\def\fueralle{\quad\mbox{f"ur alle }}
\def\fuerein{\quad\mbox{f"ur ein }}
\def\mit{\quad\mbox{mit }}
\def\falls{\quad\mbox{falls }}

\def\ed{\end{document}}

\newtheorem{theorem}{Theorem}[section]
\newtheorem{lemma}[theorem]{Lemma}
\newtheorem{corollary}[theorem]{Corollary}
\newtheorem{proposition}[theorem]{Proposition}
\newtheorem{example}[theorem]{Example}

\newenvironment{Theorem}{\begin{theorem}\sl}{\end{theorem}}
\newenvironment{Lemma}{\begin{lemma}\sl}{\end{lemma}}
\newenvironment{Corollary}{\begin{corollary}\sl}{\end{corollary}}
\newenvironment{Proposition}{\begin{proposition}\sl}{\end{proposition}}
\newenvironment{Example}{\begin{example}\rm}{\end{example}}

\section{Introduction} \label{sec: introduction}
In many applied problems and especially in biomedical studies, researchers are interested in associating
an outcome variable to several explanatory variables, typically via a generalized linear or proportional
hazards regression model. Here, the explanatory variables or predictors may be continuous, nominal or ordered.
Estimates of regression parameters can be obtained via maximizing a least-squares or (partial) likelihood function.
Especially if the number of observations is small to moderate, researchers often encounter noisy estimates
of the regression parameters, possibly leading to patterns in the regression estimates that violate the a-priori
knowledge of a factor being ordered. In order to improve accuracy of estimates and efficiency of overall tests
for associations, it is tempting to use the prior knowledge of orderings in some of the regression coefficients.

From a Bayesian perspective, receiving estimators in these type of problems is straightforward using Markov
Chain Monte Carlo approaches. Pioneered in a linear model framework by \cite{gelfand_92}, Bayesian approaches
have been proposed by \cite{robert_96, dunson_03, dunson_03_cox}. We also refer to the discussion in the latter
two papers. To use Gibbs sampling to get the ordered predictor estimate in logistic regression, \cite{holmes_06}
combine the approach in \cite{gelfand_92} with an auxiliary variable technique. Note that using \eg flat priors
on the regression coefficient vector $\ve{\beta}$ it is straightforward to show that the maximum a posteriori
estimator is equal to the constrained MLE introduced in Section~\ref{sec: general}.

Although conceptually straightforward, the implementation of these Bayesian approaches is not without fallacies.
To not only get point estimates but also assess whether parameters are equal or strictly ordered across level of
predictors, one needs to borrow from more frequentist approaches and ``isotonize'' unconstrained parameter estimates
\citep{dunson_03}. Only then one can accommodate ``flat regions'', \ie successive estimates for ordered levels
that are equal.

Although there exists vast literature on frequentist estimation subject to order restrictions \citep{robertson_88},
estimation in the specific regression model discussed here has gained surprisingly little attention
\citep{mukerjee_95}. This may be due to the fact that setting up algorithms in these type of problems
is generally difficult \citep{dunson_03}, and requires approaches that need to be adapted to specific problems,
necessitating a vast literature for numerous cases of order restricted estimation.
We mention \cite{dykstra_82, matthews_98, jamshi_04, tan_07, taylor_07}, or \cite{balabdaoui_09} discussing computation
of order restricted estimates in specific regression problems, and \cite{terlaky_98, balabdaoui_04} or
\cite{rufibach_07} for estimation of probability densities under order restrictions.
Additionally, generalizations of the pool-adjacent-violaters algorithm (PAVA) to inclusion of continuous
isotonic covariates are discussed in \cite{bacchetti_89, morton_00, ghosh_07, cheng_08}
in the context of ``additive isotonic regression''. Estimation in this type of model is usually
performed using the cyclical PAVA in connection with backfitting.
However, note that we are not in this genuinely semiparametric setting, but rather
the number of levels of an ordered factor is given a priori and remains fixed for any number of observations.


Recently, a type of algorithm, which has been around in optimization theory for some decades \citep{fletcher_87},
has gained considerable attention in the statistical literature: active set algorithms. \cite{duembgen_07} use and generalize
such an algorithm to compute a log-concave density not only from \iid but even from censored data. An algorithm
similar in spirit is the support reduction algorithm discussed in \cite{groeneboom_08}. The latter authors
apply it to the estimation of a convex density and to Gaussian deconvolution. A slight generalization
of the support reduction algorithm is used to estimate a convex-shaped hazard function in \cite{jankowski_08}.
\cite{beran_08} extend active set algorithms to the estimation of smooth bimonotone functions. They illustrate their algorithm
on regression with two ordered covariates, so also treating the example dealt with in this paper. However, \cite{beran_08}
only consider least squares or least absolute deviation estimation, and at most two ordered factors.
In this paper, we propose an algorithm for an arbitrary number of ordered factors, and we also provide a characterization of the solution.

A key feature of an active set algorithm is, that although iterative, it terminates after finitely many steps, and
that the solution is finally found via an unconstrained optimization.
This implicitly implies that, as opposed to some Bayesian approaches \citep{dunson_03},
the active set algorithm is not hurt if estimates of subsequent levels turn out to be equal.
In Section~\ref{sec: general} we show that the estimation of a regression function in generalized linear models (GLM)
under the above ordered factor restriction can be easily performed using such an active set algorithm.

\paragraph{Optimal scaling}
A reviewer drew our attention to optimal scaling, where one seeks to assign numeric values to
categorical variables in some optimal way, see e.g. \cite{breiman_85, gifi_90, hastie_90},
or applied to modeling interactions in \cite{rosmalen_09}.
In \citet[Section 2]{gifi_90} categories of the original categorical variables are replaced by
``category quantifications'', and from then on the variables are considered to be quantitative.
Note that in the approach discussed in this paper, one does not necessarily look for an optimal transformation,
but rather imposes {\it a priori knowledge} on a given ordered predictor. In the example analyzed in
Section~\ref{sec: real data} it seems plausible that a higher tumor or nodal stage is associated with a higher
risk of experiencing a second primary tumor.

\paragraph{Ordered predictors} While the treatment of quantitative and grouped predictors in regression models is
straightforward, we briefly review
alternative approaches that can be applied to deal with an ordered explanatory variable $z$. Let us assume the levels
of $z$ are coded as $1, \ldots, k$ where $k \ge 2$ and the levels are increasingly ordered, \ie $1 \le \ldots \le k$.

The most straightforward way to incorporate $z$ as a predictor is simply to ignore the information about the
groups and consider it a quantitative variable. This approach implicitly assumes that the group levels
represent a true dimension, with intervals measured between adjacent categories that correspond to the chosen
coding. If the ordinal values are arbitrarily assigned rather than actually measured, the regression coefficient
is then difficult or impossible to interpret.

Supposedly the most prevalent approach to incorporate an ordered predictor $z$ in a regression model is to introduce $k-1$ dummy variables $z_2, \ldots, z_k$
where $z_i = 1\{z = i\}, i = 2, \ldots k$. This approach ignores the additional knowledge of $z$
having ordered levels, entailing that the estimated parameters $\hat \beta_2, \ldots, \hat \beta_k$
corresponding to the above dummy variables may not be increasingly ordered. This is especially relevant
in small sample studies, where noisy estimates may confuse the proper order of dummy variable coefficients.

To simplify interpretation of models, especially when interactions are to be incorporated, researchers sometimes
resort to dichotomizing a grouped factor, \ie introducing only one dummy variable $z_1 = 1\{z \le l\}$, for some
$1\le l < k$. Here, the additional knowledge about the ordered levels is not used and may cause a substantial
loss of predictive information \citep[Section 9.1]{steyerberg_09}.

Another choice may be polynomial contrasts. One then introduces new variables $z_i = i^2\{z = i\}, i = 2,\ldots, k$. To avoid correlated estimators
$\hat \beta_i$ and therefore mutually dependent tests when doing variable selection, researchers generally
prefer to modify the design matrix in order to get orthogonal polynomial contrasts.
The function \code{as.ordered()} in \proglang{R} \citep{R} does this by default.

\cite{gertheiss_08} proposed a ridge-regression related approach to perform regression with ordered factors. Consider the
predictor $z$ with ordered categories $1, \ldots, k$ and the linear regression model
\bean
    \ve{y} &=& \beta_2 \ve{z}_2 + \ldots + \beta_{k}\ve{z}_k + \ve{\eps} \nonumber \\ \label{eq: linreg}
           &=& \mat{Z} \ve{\beta} + \ve{\eps}.
\eean For simplicity, we do not consider an intercept and only one ordered factor.
The vectors $\ve{z}_j = (1\{z_i = j\})_{i=1}^n, j = 2, \ldots, k$ are vectors of dummy variables corresponding to the levels of
$z$, $\mat{Z}$ is the $n \times (k-1)$ design matrix with the $\ve{z}_j$'s as columns, $\ve{y} \in \R^n$ is the
response and $\ve{\eps}$ an \iid noise vector where $\eps_i \sim N(0, \sigma^2)$. Note that for
reasons of identifiability, \cite{gertheiss_08} assume $\beta_1=0$ and therefore omit $\beta_1 \ve{z}_1$ in (\ref{eq: linreg}).

Instead of maximizing the original likelihood $\ell(\ve{\beta})$ over $\ve{\beta}$, \cite{gertheiss_08} instead propose
to maximize a penalized version of $\ell$:
\bean
    \ell_p(\ve{\beta}) &=& \ell(\ve{\beta}) - \lambda \sum_{j=2}^k (\beta_j - \beta_{j-1})^2 .\label{eq: pen problem}
\eean
Here, $\lambda>0$ is a tuning parameter. The solution to (\ref{eq: pen problem}) can be explicitly computed as
\bean
    \ve{\hat \beta}_{\mathrm{GT}} &=& (\mat{Z}^{\trans} \mat{Z} + \lambda \mat{\Omega})^{-1} \mat{Z}^{\trans} \ve{y} \label{eq: GT}
\eean for a fixed and specified matrix $\mat{\Omega}$. The idea is that $\ve{y}$ is assumed to change slowly for adjacent
categories, a property of $\ve{\hat \beta}_{\mathrm{GT}}$ that is ``encouraged'' by the shrinkage estimator
(\ref{eq: GT}). However, note that $\ve{\hat \beta}_{\mathrm{GT}}$ may still contain two adjacent estimates
$\beta_i, \beta_{i+1}$ such that $\beta_i < \beta_{i+1}$, a somewhat undesired feature in this setting. Furthermore,
if we choose $j=1$ as our reference level (and therefore implicitly assume that $\beta_1 = 0$), it seems reasonable to
demand for the estimated coefficients that they are all positive, what is not ensured by using (\ref{eq: GT}). Finally,
further considerations are necessary to determine the tuning parameter $\lambda$.

Consider Setting (\ref{eq: linreg}) as before. In this paper, we introduce an algorithm to solve the following problem:
Maximize $\ell$ assuming that $\beta_1=0$ and under the constraint that
\bean
0 \le \beta_2 \le  \ldots \le \beta_k, \label{eq: est pars}
\eean so that we receive non-negative and adequately ordered estimated parameters $\hat \beta_2, \ldots, \hat \beta_k$
for the factor levels. This approach is appealing since the available knowledge (or our ``prior belief'') is precisely
exploited. Furthermore, constraining the space of allowed parameters can be interpreted as regularizing the estimator,
implying higher accuracy of the constrained estimate \citep{dunson_03}.
This is especially relevant in small samples. As can be seen from (\ref{eq: est pars}),
we can estimate parameters for an ordered factor such that the constraints
$0 \le \beta_2 \le  \ldots \le \beta_k$
are enforced, unlike in \cite{gertheiss_08} where the violation of these inequalities is only penalized.
In the latter approach the violation of the
first of the above inequalities, the non-negativity constraint, is not even penalized. In addition, our estimator is fully automatic,
\ie no arbitrary choices such as the coding of levels, the determination of a cutoff to pool levels, or the selection
of a tuning parameter (such as $\lambda$ above) or bandwidth are necessary.

\paragraph{Testing in order restricted models}
There is a vast literature on likelihood ratio testing in models under linear equality and inequality constraints.
For a discussion and further references on (exact) testing under restrictions in the ordinary linear regression model
see \cite{perlman_69}, \cite{wolak_87} and \cite{shapiro_88}. \cite{silva_94} and \cite{fahrmeir_94} generalize these
results to generalized linear models, especially logistic and Cox regression. As can be seen from \eqref{eq: restLRT}
below, any likelihood ratio test (LRT) is constructed as the difference of the
likelihoods at the unrestricted and the restricted maximizer of the (partial) log-likelihood function, which entails that
one needs an algorithm to compute the restricted maximizer. \citet[Section 4]{silva_94} describes an ad-hoc approach
to find the constrained estimators. However, his algorithm is non-standard and tedious to apply \citep[p. 856]{silva_94}.
The active set algorithm described here is a general framework able to tackle general optimization problems under
constraints and therefore able to compute the restricted estimators in the above mentioned tests very efficiently.
This facilitates the application of LRTs in this type of problem.

\paragraph{Statistical inference and asymptotics}
Typically, deriving asymptotic properties of shape-constrained estimators is hard, but the starting
point in all these problems \citep{groene_01, balabdaoui_04, duembgen_08} is a characterization
of the estimator, since all the estimators are defined as maximizer of some rather involved function. The most prominent
example of a theoretical treatment of a shape constrained estimator via its characterization is the greatest convex
minorant that characterizes the estimator of a monotone density \citep{grenander_56}.
In Section~\ref{sec: characterization} we characterize the solution in our problem. Besides
being the starting point for a more thorough analysis,
a characterization also allows to check whether an algorithm actually delivers the correct solution.

\paragraph{Our contribution} We propose an active set algorithm to find estimators in GLMs with
ordered predictors. The estimators strictly comply with the constraints and are found very efficiently,
and in a finite number of steps.
For identifiability reasons, most regression approaches assume that the coefficient corresponding to the
lowest level of an ordered factor is equal to 0. Our approach ensures that all coefficients corresponding to higher
levels are in fact non-negative as well.
In addition, neither the estimator nor the proposed algorithm needs a tuning parameter.
Having an efficient algorithm at hand that provides restricted estimates
facilitates the application of LRTs to check whether ordered predictors should be
included in the model. In addition, we provide a characterization of the estimator. This serves (i) as a benchmark to
verify that the algorithm indeed delivers the maximizer, (ii) gives some insight in the structure of the estimator
and (iii) marks the starting point for a more thorough (asymptotic) analysis.

\paragraph{Organization of the paper}
A general formulation of the problem is given in Section~\ref{sec: general}.
Some examples of GLMs that illustrate our new approach are discussed in Section~\ref{sec: examples}.
A description of the active set algorithm adapted to our problem is given in Section \ref{sec: active set}.
There exist special cases of the problem that allow one to find the linear regression estimator $\ve{\hat \beta}_1$
more easily than using the active set algorithm, discussed in Section~\ref{sec: special cases}. A characterization of the solutions is
given in Section~\ref{sec: characterization}. Some indications on statistical inference are provided in
Section \ref{sec: inference}. Literature on likelihood-ratio testing to check whether an ordered factor should be included
in the model is briefly discussed in Section~\ref{sec: LRT}.
A real data example from oncology is analyzed in Section~\ref{sec: real data}. Finally, a more technical description of
the algorithm and proofs are postponed to the Appendix.


\section{Setup} \label{sec: general}

We consider the general regression problem of modeling an outcome $y \in \R$ based on some feature vector $\ve{w} \in
\R^p$. Therefore, we are given a set $(y_i,(w_{ij})_{j=1}^p)$ of observations, for $i=1,\ldots,n$. Write \bea
    \ve{y} = (y_i)_{i=1}^n \in \R^n &\text{and}& \mat{W} = (\ve{w}^{\trans}_{i \cdot})_{i=1}^n \in \R^{n \times p}
\eea where $\ve{w}_{i \cdot} = (w_{ij})_{j=1}^p$, $i=1,\ldots,n$. The predictors are denoted by
$\mat{w}_{\cdot j} = (w_{ij})_{i=1}^n$ for $j = 1, \ldots, p$. Throughout the exposition, $n$ and $p$ are considered
to be fixed.

In general, for given $\ve{y}$ and $\mat{W}$, we seek to maximize a real--valued concave criterion function \bea
    L = L(\ve{y}, \mat{W}, \ve{\beta}): \R^{n} \times \R^{n \times p} \times \R^p \to \R
\eea over $\ve{\beta} \in \R^p$, yielding an estimated parameter vector $\ve{\hat \beta} \in \R^p$. Note that
to define our estimator and to derive the characterization in Section~\ref{sec: characterization}, a model needs no
further specification that goes beyond the function $L$.
Ordinary, \ie unordered, factors are assumed to be already coded as dummy variables, so they are considered
quantitative. If an intercept is to be taken into the model, we simply assume it to be a quantitative
variable of all $1$'s.
Let $c$ denote the number of quantitative predictors and suppose that the last $f$ predictors $\mat{w}_{\cdot j}$,
$j = c + 1, \ldots, p$ are ordered factors, each with $k_j$ levels (so $c = p-f$). Furthermore, the coding is
assumed such that $w_{ij} \in \{1, \ldots, k_j\}, \, i = 1, \ldots, n$, where a higher number corresponds to a
``higher'' level of the ordered factor $\mat{w}_{\cdot j}$.
Introduce the sets of indices $\JJcp = \{c+1,\ldots,p\}$ and $\LLj = \{2, \ldots, k_j \}$ for $j \in \JJcp$.
Clearly, the case $c=0$ (no quantitative variables in the model) is not excluded. However, we assume to have at least one
ordered factor, \ie $f \ge 1$ which immediately implies $p\ge 1$.
In order to respect the ordinal character of each of the factors $\mat{w}_{\cdot j}$ we estimate $\ve{\beta}$
based on a new data matrix $\mat{X} \in \R^d$. This latter matrix is obtained via modifying the original data matrix $\mat{W}$
by adding
\bea
    f \cdot \Bl(\Bl(\sum_{j=c+1}^p k_j\Br) - (p - c)\Br)
\eea dummy variables for the levels $\ge 2$ of the ordered factors. We then constrain
optimization of the updated functional $L = L(\ve{y}, \mat{X}, \ve{\beta})$ to the constrained space of
parameters
\bean
    \BB(c, \, p, \, \ve{k}) &=& \Bl\{\ve{\beta} \in \R^d \ : \ \beta_{j, \, 2} \ge 0, \ \beta_{j, \, l+1} - \beta_{j,\, l} \ge 0, \  \ l \in \LLj, j \in \JJcp\Br\}. \hspace{0.5cm}\label{eq: general B}
\eean Here, $\beta_{j,\, l}$ is the coefficient of the dummy variable corresponding to the level $l$ of the
$j$-th ordered factor, and $\ve{k} = ((0)_{i=1}^c, k_{c+1},\ldots,k_p) \in \R^p$.
For ease of notation, we define $\BB = \BB(c, \, p, \, \ve{k})$.
Constraining estimation to $\BB$ ensures that the estimated parameter corresponding to a
``higher'' level of an ordered factor is at least as large as those of ``lower'' levels and all estimated
parameters are non-negative.
Note that our approach also adds something new if we have an ordered factor with only two levels (note that we always
lose the level attributed to the baseline), namely that $\beta_{j, 2} \ge 0$ for this ordered factor.


\section{Examples} \label{sec: examples}

We briefly specify the GLMs we provide algorithms for. Extensions to other criterion functions are straightforward.

\paragraph{Linear regression}
Here, $\ve{y} \in \R^n$ and we estimate $\ve{\beta}$ via
maximizing the criterion function $\ell_{n,1}$ over all $\ve{\beta} \in \BB$. This latter function is defined as
\bea
    \ell_{n,1} (\ve{\beta}) &=& - \sumn ( y_i - \ve{x}_{i \cdot}^{\trans} \ve{\beta})^2.
\eea Here, $\ve{x}_{i \cdot}$ denotes the $i$-th row vector of $\mat{X}$.
We emphasize that given $L$ 
there is no need to further specify a model for the data.

\paragraph{Logistic regression} In this case, $\ve{y} \in \{0, 1\}^n$. Using maximum likelihood estimation (MLE)
we obtain the log--likelihood function
\bea
    \ell_{n,2}(\ve{\beta}) &=& - \sumn \Bl(-y_i \ve{x}_{i \cdot}^{\trans} \ve{\beta} + \log (1+\exp(\ve{x}_{i \cdot}^{\trans} \ve{\beta}))\Br).
\eea

\paragraph{Cox regression} Here, we have observations
$(T_i, C_i, \delta_i, \ve{x}_{i \cdot})$ for $i=1, \ldots, n$. Clearly, $T_i$ are the failure times (possibly
unobserved), $C_i$ the censoring times, $\delta_i = 1\{$event has happened$\}$ and $\ve{x}_{i \cdot}$ is the
feature vector as before. If we introduce the observed time $V_i =\min\{T_j , C_j\}$ for each unit, let
\bea
    R_i &=& \{j \, : \, V_j \ge T_i\}
\eea denote the number of individuals at risk after time $T_i$, $i = 1, \ldots, n$. The partial
likelihood according to \cite{cox_72} is then
\bea
     \prod_{i=1}^n \Bl[\frac{\exp(\ve{x}_{i \cdot}^{\trans}\ve{\beta} )}{\sum_{k \in R_i}\exp(\ve{x}_k^{\trans}\ve{\beta})} \Br]^{\delta_i}.
\eea Introducing $\alpha_i = \exp (\ve{x}_{i \cdot}^{\trans}\ve{\beta})$ for $i = 1, \ldots, n$ and
letting $t_1 < \ldots < t_D$ be the observed (assumed to be distinct, for simplicity) event times, we then easily deduce the
log-likelihood function: \bea
    \ell_{n,3}(\ve{\beta}) &=& \sumn \alpha_i - \sumn \delta_i \log\Bl(\sum_{k \in R_i}\alpha_k \Br) \\
                  &=& \sum_{s=1}^D \alpha_{(s)} -\sum_{s=1}^D  \log\Bl(\sum_{k \in R_s}\alpha_k\Br)
\eea where $\alpha_{(s)}$ is the above expression belonging to the $s$-th failure time, $s=1,\ldots,D$.

\paragraph{Properties of the maximization problems} Let us introduce the constrained
\bean
    \ve{\hat \beta}_i &:=& \underset{{\ve{\beta} \, \in \, \BB}}{\text{maximize }} \ell_{n,i}(\ve{\beta}) , \ i = 1, 2, 3 \label{eq: general constr opt problem}
\eean and the unconstrained
\bea
    \ve{\hat \eta}_i &:=& \underset{{\ve{\beta} \, \in \, \R^d}}{\text{maximize }} \ell_{n,i}(\ve{\beta}) , \ i = 1, 2, 3 \label{eq: general unconstr opt problem}
\eea maximizers. The conditions on fixed response $\ve{y}$ and design matrix $\mat{X}$ under which
$\ve{\hat \eta}_i$ exist and are unique in logistic regression are well studied \citep{albert_84, santner_86}.
\cite{silva_86} specify necessary and sufficient conditions for the MLE to exist in logistic and
Cox regression. Since the set $\BB$ is a closed convex cone, the estimators $\ve{\hat \beta}_i$ exist and are unique
for $i = 1, 2, 3$ at least under the same conditions as those for $\ve{\hat \eta}_i$.
Conditions for consistency and asymptotic normality of MLEs in GLMs are provided in \cite{fahrmeir_85}.
In this paper we assume that our design matrix $\mat{X}$ is such that $\ell_{n, i}$ is concave and coercive
for $i = 1, 2, 3$.

\section{Active set algorithm to compute $\ve{\hat \beta}_i$} \label{sec: active set}
In \cite{fletcher_87} an active set algorithm is described, a useful tool for constrained optimization
problems. In connection with likelihood ratio tests (see Section~\ref{sec: LRT}) we came across \cite{silva_94}.
In Section~4 of this latter paper, it seems as if a version of the active set algorithm is described. However,
instead of directly computing the ``active set'' in each iteration (see below), a crude and
computationally expensive ``all-subset search'' is proposed. In the context of mixture models, the algorithm discussed
by \cite{groeneboom_08} can also be interpreted as a variant of an active set algorithm.

In Section 3 of \cite{duembgen_07} the general principle of active set algorithms is described in detail, complemented
by a discussion of its validity. Here, we
therefore limit ourselves to the discussion of the main features and points relevant for the application of the
active set algorithm to find the $\ve{\hat \beta}_i$'s. We briefly sketch the idea of an active set algorithm,
and refer to \ref{sec: details algo} for a detailed technical exposition of the algorithm for the problem
treated here.

Let $q$ denote the number of constraints that compose $\BB$, $\ell$ the function to be maximized
and $\ve{\beta}$ its maximizer,
see Section~\ref{sec: examples}. Define for any index set $A \subseteq \{1, \ldots, q\}$
the linear subspace
\bea
    \VV(A) &=& \Bl\{\ve{\beta} \in \R^d \ : \ -\beta_{j, \, l} + \beta_{j, \, l - 1} 1\{l \ge 3\} = 0, \ \text{ for all } j, \, l \text{ such that } \phi(j, \, l)\in A\Br\}.
\eea The function $\phi$ maps the indices $j$ and $l$ of the dummy variables forming the ordered factors to
the number of constraining inequalities, see \eqref{eq: def phi} in \ref{sec: details algo}. The crucial assumption for an active set algorithm is
that we have another algorithm available that for any $A \subseteq \{1, \ldots, q\}$ (efficiently) computes
\bea
    \ve{ \til\beta}(A) &=& \argmax_{\ve{\beta} \, \in \, \VV(A)} \ell(\ve{\beta}), \label{eq: max on subspace}
\eea provided that $\VV(A) \cap \{\ve{\beta} \ : \ \ell(\ve{\beta}) > -\infty\} \ne 0$.
Subspaces of the parameter space are considered when violations of the initial constraints appear in the algorithm.
In this case, the active set algorithm varies $A$ in a deterministic way, until finally $\ve{\til\beta}(A) = \ve{\hat \beta}$.
In order to tailor an active set algorithm to a specific
problem, the above maximization on a subspace is crucial. In our regression with ordered covariates setting, we show
in \ref{sec: details algo} (see Table~\ref{table: possible violations}) that three types of subspaces
have to be dealt with, depending on the specific violation that occurs.

It is important to realize that by design, the main routine of an
active-set algorithm does not need a stopping criterion as \eg Newton-type algorithms. Once the algorithm has identified the set $A$ that corresponds
to the solution $\ve{\hat \beta}$, it performs
an unrestricted maximization (here, a stopping criterion may be necessary), which at least in the
linear, logistic and Cox regression examples is unproblematic.
Verification that a given $\ve{\hat \beta}$ is the maximizer can be done by means of Theorem 3.1 in \cite{duembgen_07}.
Additionally, since there are only finitely many subsets of $A$, the algorithm terminates after finitely many steps.

\section{Special case: An almost explicit solution} \label{sec: special cases}
To be able to state the following results concisely, let us introduce for every ordered factor
$j \in \JJcp$ the set of indices where the equality constraint $\hat \beta_{j, l} \ge 0$ is active:
\bea
    \ZZ_{j}(\ve{\hat \beta}_i) &=& \Bl\{l \in \{2, \ldots, k_j\} \, : \, \hat \beta_{j, l} = 0\Br\}
\eea for all $j \in \JJcp$.

In this section, we restrict our attention to the case of linear regression with only one ordinal predictor. If in addition
$\ZZ_1(\ve{\hat \beta}_1) = \emptyset$, that means the constrained estimator has only strictly positive
entries anyway, then $\ell_{n,1}$ simplifies such that $\ve{\hat \beta}_1$ can be found via solving (\ref{eq: min prob pava}).
\begin{lemma}\label{lem: simple case PAVA} If $c = 0$, $f=1$ and $\ZZ_1(\ve{\hat \beta}_1) = \emptyset$, the estimator
$\ve{\hat \beta}_1$ is
\bean
    \ve{\hat \beta}_1 &=& \argmax_{\beta_2 \ge \ldots \ge \beta_{k_1}} \ell_{n,1}(\ve{\beta}) \nonumber \\
    &=& \argmin_{\beta_2 \le \ldots \le \beta_{k_1}} \sum_{j=2}^{k_1} N_j (\beta_j - m_j)^2 \label{eq: min prob pava}
\eean where for $l \in \LL_1$ \bea
    N_l = \sumn 1\{x_{il} = 1\} &\text{ and }& m_l = {N_l}^{-1}\sum_{i:x_{il} = 1} y_i.
\eea
\end{lemma}
\noindent The proof of this lemma is postponed to \ref{sec: proofs}.


The solution to \eqref{eq: min prob pava} can easily be computed
using the PAVA \citep{barlow_72, robertson_88}. This latter algorithm performs
at most $n-1$ iterations until the vector $\ve{\hat \beta}_1$ is found.

One of the initial motivations to analyze regression with ordered predictors, and the reason why
we included this very specific example, was to see whether this simple and appealing structure can be carried forward to
the more general problem of more ordered factors and additional quantitative variables.
However, since (i) we were not able to construct a generalized PAVA algorithm that solves our problem and
(ii) we are not only interested in the least-squares problem but also treat GLMs, we switched to an active set algorithm.

\section{Characterization of the solution} \label{sec: characterization}
There are two main purposes of providing a characterization of the estimator $\ve{\beta}$:
(i)~knowing the structure of the maximizer of $\ell$ allows one to cross-check the validity of the proposed active-set
algorithm and to check whether it has found the
correct maximizer of $\ell$. (ii)~It is well-known that in such constrained estimation problems, the key to deriving
asymptotic properties of the estimator such as consistency or rate of convergence is a characterization in terms
of directional derivatives, see the discussion in Section~\ref{sec: introduction}.

To be able to state the following theorem properly, we introduce the function
$\psi: \bigl\{(c+1) \times \LL_{c+1},\ldots,p \times \LL_p \bigr\}\to \{1, \ldots, d\}$ that
maps the original indices $(j, \, l)$ to the column number of the respective dummy variable in $\mat{X}$,
or equivalently, to the index $i$ that corresponds to the entry of the vector $\ve{\beta} \in \BB(c, p, \ve{k})$
that corresponds to $\ve{\beta}_{j, \, l}$. Specifically, this function is for any $j \in \JJcp$
and $l \in \LLj$,
\bean
    \psi(j, \, l) &=& c + \Bl(\sum_{h=c+1}^{j} k_{h-1}\Br)+(l-1)-(j-c-1) \nonumber \\
    &=& 2c + \Bl(\sum_{h=c+1}^{j} k_{h-1}\Br)+l-j \label{eq: def psi}.
\eean By $\psi^{-1}$ we denote the inverse of this function, i.e. the function that maps the position $i$ of the entry
of $\ve{\beta}$ to the indices $j$ and $l$. Now, for each $j \in \JJcp$ let
$\ve{h}_j$ be the vector of distinctive strictly positive values of $(\ve{\beta}_j)_{j \in \LLj}$ for every $j \in \JJcp$
and any $\ve{\beta} \in \BB$. Using these definitions we split any vector $\ve{\beta} \in \BB$ into the following blocks:

\bigskip

\begin{tabular}{lcll}
$B_1(\ve{\beta})$         & \ = \ & $\{i \ : \ i = 1,\ldots, c \}$ & (coefficients of quantitative variables), \vspace*{0.2cm} \\
$B_{2, j}(\ve{\beta})$    &  \ = \ & $\{i \ : \beta_{\psi^{-1}(i)} = 0\ \text{ and } (\psi^{-1}(i))_1 = j\}$, &  \vspace*{0.2cm} \\    
$B_{3, j, u}(\ve{\beta})$ &\ = \ & \multicolumn{2}{l}{\{all indices $i$ s.t. $\beta_{\psi^{-1}(i)} = h_{j, u}$ for $(\psi^{-1}(i))_1 = j$\}, }
\end{tabular}

\bigskip

\noindent where $u = 1,\ldots, |\ve{h}_j|$ for each $j$. Here, $|.|$ denotes the dimension of a vector $\ve{a}$ or the number of elements in a set. Note that
$|B_1|+|\cup_j B_{2,j}|+|\cup_{j, u} B_{3,j, u}|= d$.
Using these blocks, we are now able to formulate the characterization of the solution.

\begin{theorem} \label{lem: characterization}
An arbitrary vector $\ve{\hat \gam} \in \BB(c, p, \ve{k})$ maximizes the concave function $\ell$ if and only if it
fulfills the following conditions:
\bean
    \Bl(\nabla \ell(\ve{\hat \gamma})\Br)_s &=& 0 \ \text{ for all } \ s \in B_1(\ve{\hat \gam})  \label{eq: charac1} \\
    \sum_{s = \min B_{3, j, u}(\ve{\hat \gam})}^t \Bl(\nabla \ell(\ve{\hat \gamma})\Br)_s &\ge& 0 \, , \ \text{ for all } t \in B_{3, j, u}(\ve{\hat \gam}), \, u = 1, \ldots, |\ve{h}_j| \text{ and } j \in \JJcp, \label{eq: charac2}  \\
    \sum_{s = t}^{\max B_{3, j, u}(\ve{\hat \gam})} \Bl(\nabla \ell(\ve{\hat \gamma})\Br)_s &\le& 0 \, , \ \text{ for all } t \in B_{3, j, u}(\ve{\hat \gam}), \, u = 1, \ldots, |\ve{h}_j| \text{ and } j \in \JJcp. \label{eq: charac3}
\eean
\end{theorem}

\noindent Note that the entries of the gradient at the active constraints $\beta_i$, $i \in B_{2, j}$, are not needed to
characterize the solution since $\ve{\hat \gamma}$ always equals 0 at these positions.
Furthermore, the theorem immediately implies
\bean
    \sum_{s \in B_{3, j, u}} \Bl(\nabla \ell(\ve{\hat \gamma})\Br)_s & = & 0  \label{eq: charac4}
\eean for $u = 1,\ldots, |\ve{h}_j|$ and $j \in \LLj$.

To illustrate Theorem~\ref{lem: characterization}, consider the following example: For $n = 200$ observations
we generated a dataset with standard normally distributed errors, three quantitative variables, one
(unordered) factor (with three levels) and one ordered factor (with eight levels). The model we stipulated to generate
the response $\ve{y}$ was
\bea
    y_i &=& 2 q_{1i} - 3 q_{2i} + 0 q_{3i} + 0 f_{1i} + f_{2i} + f_{3i} + 0 o_{1i} + 0 o_{2i} + 2 o_{3i} + 2 o_{4i} + 2 o_{5i} + 2 o_{6i} + 5 o_{7i}
    + 5 o_{8i} + \epsilon_i
\eea where $q_{ji} \sim N(1, 2)$ for $j = 1, 2, 3$ and $i=1, \ldots, n = 200$, each level of any factor (whether ordered or unordered)
has the same number of observations and these are randomly allocated to the observations.
Finally, $\epsilon_i \sim N(0, 4)$ for $i = 1, \ldots, n$.
The resulting (constrained) linear regression estimates are given in Table~\ref{tab: example1}.
Note that for comparison we also added columns for the estimator $\ve{\hat \rho}_1$ which is computed similarly
to $\ve{\hat \beta}_1$, but without the positivity restriction $\beta_{6, 2} \ge 0$. For this estimator,
a characterization similar to that in Theorem~\ref{lem: characterization} can be given using exactly the same
approach.

\begin{table}
\begin{center}
\renewcommand{\baselinestretch}{1.2}
{\footnotesize
\begin{tabular}{cccrrrrrrrr}
 Var & Level & $\ve{\beta}$ & $\ve{\hat \eta}_1$ & $\nabla \ve{\hat \eta}_1$ & $\ve{\hat \rho}_1$ & $\nabla \ve{\hat \rho}_1$ & $\nabla_\uparrow \ve{\hat \rho}_1$ & $\ve{\hat \beta}_1$ & $\nabla \ve{\hat \beta}_1$ & $\nabla_\uparrow \ve{\hat \beta}_1$ \\
  \hline
quant &  & 2 & 2.13 & 0 & 2.12 & 0 & 0 & 2.08 & 0 & 0 \\
  quant &  & -3 & -2.95 & 0 & -2.94 & 0 & 0 & -2.96 & 0 & 0 \\
  quant &  & 0 & 0.19 & 0 & 0.18 & 0 & 0 & 0.17 & 0 & 0 \\
   \hline
fact1 & 2 & 1 & 1.06 & 0 & 1.08 & 0 & 0 & 0.88 & 0 & 0 \\
  fact1 & 3 & 1 & 1.53 & 0 & 1.41 & 0 & 0 & 1.23 & 0 & 0 \\
   \hline
ord1 & 2 & 0 & -0.85 & 0 & -0.78 & 0 & 0 & 0 & -26.86 & -26.86 \\
  ord1 & 3 & 2 & 3.55 & 0 & 1.99 & 79.8 & 79.8 & 2.19 & 79.23 & 52.38 \\
  ord1 & 4 & 2 & 1.67 & 0 & 1.99 & -13.57 & 66.23 & 2.19 & -12.66 & 39.71 \\
  ord1 & 5 & 2 & 0.60 & 0 & 1.99 & -65.85 & 0.39 & 2.19 & -65.3 & -25.59 \\
  ord1 & 6 & 2 & 1.94 & 0 & 1.99 & -0.39 & 0 & 2.19 & -1.27 & -26.86 \\
  ord1 & 7 & 5 & 4.41 & 0 & 4.47 & 0 & 0 & 4.65 & 0 & -26.86 \\
  ord1 & 8 & 5 & 4.55 & 0 & 4.60 & 0 & 0 & 4.79 & 0 & -26.86 \\
  \hline
\end{tabular}
}
\end{center}
\vspace*{-0.5cm}
\caption{Estimators, gradients and cumulative sum of gradients for Example 1.
$\ve{\hat \eta}_1$ is the unconstrained estimate, $\ve{\hat \rho}_1$ the constrained version without
the non-negativity restriction, and $\ve{\hat \beta}_1$ the restricted and non-negative estimate.}
 \label{tab: example1}
\end{table}

In this example, we get the following quantities: $p=6$, $f=1$, $c=5$, $d=12$,
$\JJ_{5, 6}=\{6\}$, $\LL_6=\{2, \ldots, 8\}$,
$k_6 = 8$, and finally
\bea
    \BB(5, 6, 8) &=& \Bl\{\ve{\beta} \in \R^{12} \ : \ \beta_{6, \, 2} \ge 0, \ \beta_{6, \, l+1} \ge \beta_{6,\, l} , \  \ l \in \{2, \ldots, 7\}\Br\}.
\eea
The notation $\nabla_\uparrow \ve{v}$ in Table~\ref{tab: example1} is shorthand for the cumulative sum of any vector
$\ve{v} \in \R^d$:
\bea
\nabla_\uparrow \ve{v} &=& (\sum_{i=1}^k v_i)_{k=1}^d.
\eea The values of the least-squares criterion function for the three estimates are
\bea
\ell_{n,1}(\ve{\hat \eta}_1) \ = \ -2964.8 \hspace*{1cm}
\ell_{n,1}(\ve{\hat \rho}_1) \ = \ -3074.8\hspace*{1cm}
\ell_{n,1}(\ve{\hat \beta}_1) \ = \ -3085.3
\eea and $\ZZ_6(\ve{\hat \beta}_1)=\{2\}$.

Let us now illustrate Theorem~\ref{lem: characterization}.
For either quantitative variables or dummy variables corresponding to unordered factors
(which in our context are conceptually equivalent), the respective entry of the gradient $\nabla \ve{\hat \beta}_1$ is always 0. As for the ordered factor,
for the entries where the positivity constraint is active (i.e. the elements in $\ZZ_6(\ve{\hat \beta}_1)$), the
gradient has a value which is not used (and not necessary) for a characterization of $\ve{\hat \beta}_1$.
The sets defined above are for the simulated example:

\bigskip

\begin{center}
\begin{tabular}{lcllcllcl}
$B_{1,6}$     & = & $\{1, \ldots, 5\}$ & $B_{2,6}$    & = & $\{6\}$  & $B_{3, 6, 1}$ & = & $\{7, \ldots, 10\}$ \\
$B_{3, 6, 2}$ & = & $\{11\}$           & $B_{3,6, 3}$ & = & $\{12\}$ & $\ve{h}_6$    & = & $(2.19, 4.65, 4.79).$
\end{tabular}
\end{center}

\bigskip

\noindent These sets then yield the following inequalities, according to (\ref{eq: charac2}) and (\ref{eq: charac3}):
\bea
\Bl(\nabla \ell_{n,1}(\ve{\hat \beta}_1)\Br)_s \ = \ 0 \ \text{ for } \ s \in \{1, \ldots, 5\} && \Bl(\nabla \ell_{n,1}(\ve{\hat \beta}_1)\Br)_s \ = \ 0 \ \text{ for } \ s = 6\\
\sum_{s=7}^{t} \Bl(\nabla \ell_{n,1}(\ve{\hat \beta}_1)\Br)_s \ \ge \ 0 \ \text{ for } \ t \in \{7, \ldots, 10\} &&
\sum_{s=t}^{10} \Bl(\nabla \ell_{n,1}(\ve{\hat \beta}_1)\Br)_s \ \le \ 0 \ \text{ for } \ t \in \{7, \ldots, 10\} \\
\Bl(\nabla \ell_{n,1}(\ve{\hat \beta}_1)\Br)_s \ = \ 0 \ \text{ for } \ s \in \{11, 12\}. &&
\eea

\section{Statistical inference} \label{sec: inference}

Having shown how to compute estimators $\ve{\hat \beta}_i$ for $i = 1,2, 3$, the question arises how to perform
(frequentist) statistical inference in these models.
Deriving consistency, rate of convergence and limiting distributions for estimators similar to $\ve{\hat \beta}_i$ under
standard assumptions is known to be non-trivial.
It is therefore not clear how to construct e.g. confidence intervals for our estimated parameters
of the ordered factor.
By using the characterization given in Section~\ref{sec: characterization}, one should be able
to derive rates of convergence and even the limiting distribution of $\ve{\hat \beta}$ as $n \to \infty$
in a suitably specified model, thereby generalizing the results of \cite{brunk_70} and \cite{wright_81} for isotonic regression
to our more general setting.
This, together with a generalization of the likelihood ratio tests introduced in Section~\ref{sec: LRT} to an arbitrary
number of ordered factors, is subject to ongoing research.

Note that bootstrap is not without fallacies in these type of models, see \cite{kosorok_09} and \cite{sen_09}.
\section{Testing for the presence of constraints} \label{sec: LRT}
There is a vast literature on likelihood ratio testing in models under linear equality and inequality constraints.
For a discussion and further references on (exact) testing under restrictions in the ordinary linear regression model
see \cite{perlman_69}, \cite{wolak_87} and \cite{shapiro_88}. \cite{silva_94} and \cite{fahrmeir_94} generalize these results to
generalized linear models, especially logistic and Cox regression.
Suppose a researcher wants to test the following hypotheses:
\bean
    H_0 \ : \ \beta_{c+1, 2} \ =  \ \ldots \ = \ \beta_{c+1, k_{c+1}} \ = \ 0 \quad \text{ vs. } \quad
    H_1 \ : \ \ve{\beta} \in \BB(c, c+1, k_{c+1}). \label{eq: hypo LRT}
\eean
Note that the estimator under $H_0$ can be computed via an unrestricted maximization. It corresponds to
a maximization using a modified design matrix $\mat{X}$ with the columns $\psi(c+1, 2), \ldots, \psi(c+1, k_{c+1})$ omitted.
Since under $H_0$ we need to consider an unrestricted estimator, we have to constrain attention either to (i) only one
ordered factor or (ii) a test of inclusion of all ordered factors against their entire exclusion from the model.
The potential influence of the additional ordered factor(s) on the response is assessed with $H_1$.
In notation similar to \cite{silva_94}, the above hypotheses translate to
\bea
    H_0 \ : \ \mat{R} \ve{\beta} \ = \ \ve{0} \quad \text{ vs. } \quad H_1 \ : \ \mat{R}_2 \ve{\beta} \ \ge \ \ve{0},
\eea where here $\mat{R} = \mat{R}_2$ is the $k_{c+1} \times d$ matrix chosen such that
\bea
    \mat{R}_2 \ve{\beta} &=& \Bl((0)_{i=1}^c, \, \beta_2, \, \beta_3 - \beta_2, \, \ldots, \, \beta_{k_{c+1}}-\beta_{k_{c+1}-1} \Br)^{\trans}.
\eea Following the development in \cite{silva_94}, the likelihood ratio test statistic to test the Hypotheses
(\ref{eq: hypo LRT}) is defined as
\bean
    T_\mathrm{LR} &=& 2 \Bl(\ell(\ve{\hat \beta})-\ell(\ve{\hat \eta})\Br). \label{eq: restLRT}
\eean The distribution of $T_\mathrm{LR}$ is a mixture of $\chi^2$ distributions.
The weights are in principle fully specified, however, in general hard to compute
\citep{wolak_87}. As a remedy, one can either use exact Monte Carlo weights \citep{wolak_87}
or bounds on the $p$-value for the above test \citep[Proposition~1]{silva_94}.
%

As can be seen from \eqref{eq: restLRT} any LRT is constructed as the difference of the likelihoods
at the unrestricted and the restricted maximizer of the (partial) log-likelihood function, which entails that
one needs an algorithm to compute the restricted maximizer. \citet[Section 4]{silva_94} describes an ad-hoc approach
to find constrained estimators. However, his algorithm is non-standard and tedious to apply \citep[p. 856]{silva_94}.
The active set algorithm described here is a general framework able to tackle general optimization problems under
constraints and able to compute the restricted estimators in the above mentioned tests very efficiently.


\section{A real data example} \label{sec: real data}

We illustrate our new algorithm using a data set from oncology, initially analyzed in \cite{taussky_05}.
The goal of the study was to assess the impact of treatment- and patient-related factors on the risk of developing
a second primary tumor (SPT) of the upper aerodigestive tract within three years after initial therapy,
in head-and-neck cancer patients.
For a subset of 231 patients that had either been observed at least three years without SPT or experienced
an SPT before three years, the endpoint
\bea
    \mathrm{SPT}_3 &=& 1\{\text{The patient experienced a SPT at 3 years or before}\}
\eea was defined and modeled using multiple logistic regression. The explanatory variables are described in
Table~\ref{tab: SPT example}.

\begin{table}
{\footnotesize
\begin{center}
\begin{tabular}{lll}
Variable &   Type   &   Levels (first mentioned = baseline) \\
\vspace*{-0.3cm} \\
\hline
\vspace*{-0.3cm} \\
Intercept (inter)                        & constant                                  & $-$ \\
Age (age)                                & continuous (standardized) \hspace*{0.2cm} & $-$ \\
Treatment (tmt)                          & factor                                    & Chemotherapy (CT) yes, CT no \\
Radiotherapy (rt)                        & factor                                    & concomitant boost (CB), hyperfractionation (HF) \\
Sex (sex)                                & factor                                    & female, male \\
Tumor stage (t)                          & ordered factor                            & $1<2<3<4$ \\
Nodal stage (n)                          & ordered factor                            & $1<2<3<4<5<6$ \\
Performance status (ps)  \hspace*{0.2cm} & ordered factor                            & $1<2<$``stage greater than 2''
\end{tabular}
\end{center}
\vspace*{-0.5cm}
\caption{Explanatory variables in real data example.} \label{tab: SPT example}
}
\end{table}

\begin{figure}[!h]
\begin{center}
\centerline{\includegraphics[width=14cm]{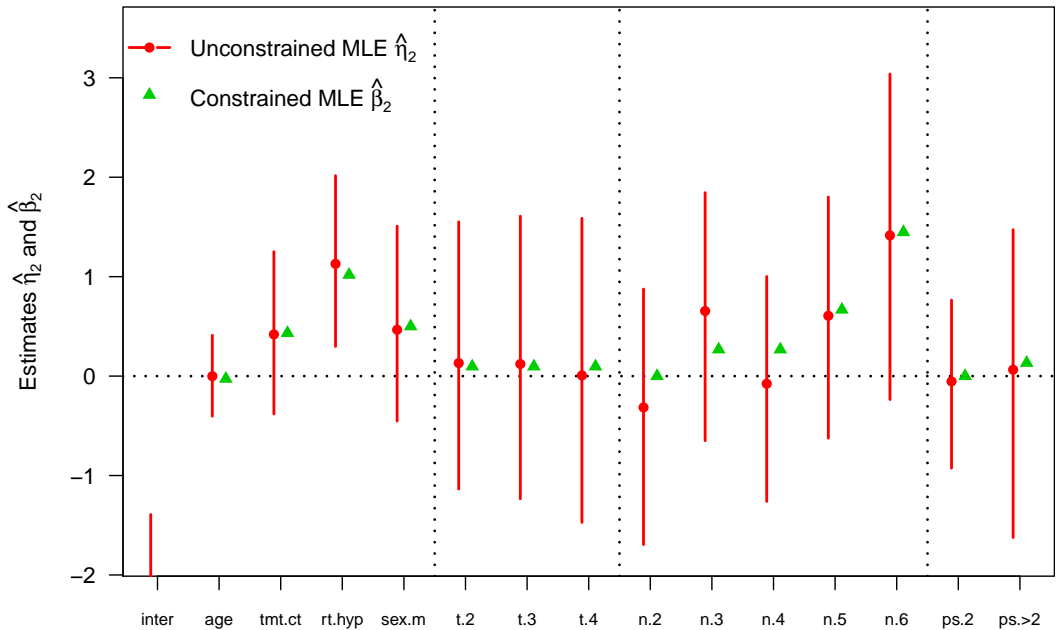}}
\end{center}
\vspace*{-1cm}
\caption{Estimates and confidence intervals for SPT example.} \label{fig: SPT example}
\end{figure}

Researchers assume in general that higher tumor stage, nodal stage, and performance status correspond to a higher
risk of experiencing a SPT. It seems therefore appropriate to use our constrained estimator in this setting.
In Figure~\ref{fig: SPT example}, the unconstrained and constrained estimators $\ve{\hat \eta}_2$ and $\ve{\hat \beta}_2$
are displayed (dot and triangle, respectively) as well as profile likelihood confidence intervals for $\ve{\hat \eta}$
($\alpha = 0.05$).
Values of the likelihoods were $\ell_{n,2}(\ve{\hat \eta}_2) = -101.6$ and $\ell_{n,2}(\ve{\hat \beta}_2) = -102.2$.

Estimates for quantitative predictors, \ie those for age, treatment, radiotherapy and sex turned out to be very similar
for $\ve{\hat \eta}_2$ and $\ve{\hat \beta}_2$. On the other hand, the ``prior belief'' or assumption of non-negative
and increasing estimates for the levels of the ordered factors tumor and nodal stage and performance status was
violated by the unconstrained estimator $\ve{\hat \eta}_2$ and ``corrected'' by $\ve{\hat \beta}_2$.

The original analysis in \cite{taussky_05} focused on identifying factors that influence the occurrence of SPT.
Variables were not taken into account as ordered factors, but were dichotomized. For comparison, we also computed
the restricted and unrestricted estimates in this setting, see Table~\ref{tab: SPT example original}.
It turns out that parameter estimates and corresponding odds ratios (OR) for the two approaches were similar, except for the nodal status. Note that
the effect of tumor stage is reversed, compared to the case where we consider all factor levels (and do not only
dichotomize), compare Figure~\ref{fig: SPT example}.

\clearpage

\begin{table}
{\footnotesize
\begin{center}
\begin{tabular}{lllrrrr}
Variable    &      Type     &      Levels   &    $\ve{\hat \eta}_1$     &    OR     &    $\ve{\hat \beta}_1$    &    OR \\
\vspace*{-0.3cm} \\
\hline
\vspace*{-0.3cm} \\
Intercept                           &    constant                           &   &-2.56   &      &   -2.72   &       \\
Age                                 &    factor                             &    $\le57$, $>57$         &   -0.19   &   0.83    &   -0.20   &   0.82    \\
Treatment                           &    factor                             &    CT yes, CT no      &   0.41    &   1.51    &   0.42    &   1.53    \\
Radiotherapy                        &    factor                             &    CB, HF                     &   1.03    &   2.81    &   0.99    &   2.70    \\
Sex                                 &    factor                             &    female, male               &   0.51    &   1.67    &   0.49    &   1.63    \\
Tumor stage                         &    ordered factor                     &    $1$, $>1$  &   -0.21   &   0.81    &   0.00    &   1.00    \\
Nodal stage                         &    ordered factor                     &    $0$, $>0$  &   0.26    &   1.29    &   0.27    &   1.31    \\
Performance status \hspace*{0.2cm}  &    ordered factor \hspace*{0.2cm}     &    $0$, $>0$  &   0.37    &   1.44
&   0.40    &   1.49
\end{tabular}
\end{center}
\vspace*{-0.5cm}
\caption{Explanatory variables in real data example, dichotomized variables as in original paper.}
\label{tab: SPT example original}
}
\end{table}

\section{Extensions} \label{sec: extensions}
It is straightforward to generalize the set $\BB(c, p, \ve{k})$ to
\bea
    \BB'(c, \, p, \, \ve{k}, \ve{r}) &=& \Bl\{\ve{\beta} \in \R^d \ : \ \beta_{j, \, 2} \ge r_{j, 2}, \ \beta_{j, \, l+1} - \beta_{j,\, l} \ge r_{l+1}, \  \ l \in \LL_j \setminus \{k_j\}, j \in \JJcp\Br\} \hspace{0.5cm}\label{eq: B, general R}
\eea for arbitrary real numbers $r_{j, \, l}$. Using such a more general parameter space could be beneficial in connection
with finding the minimum effective dose in dose-response models. The dose levels would then take the role of an ordered
factor \citep{wang_07}. Our new approach easily allows us to incorporate further predictors of any
of the three types described in the introduction to model the response.

Modeling a factor with {\it decreasing} levels can be achieved by reversing the coding of the corresponding
ordered factor, using the algorithm under the constraint of increasing levels and finally re-reversing the order
of the estimates in the vector $\ve{\hat \beta}$.
Using this approach, it is straightforward to find the solutions for all combinations of
possible orderings of, say, three ordered factors. By computing the value of the criterion function for all these
resulting coefficient vectors, one can find the one with the lowest criterion value, an approach related
to finding a global maximum in the criterion function described in \cite{kooij_06}.

Generalizations to further criterion functions, such as other GLMs or least absolute deviation regression with ordered
covariates, are straightforward. As for the latter problem, we suggest smoothly approximating the not everywhere
differentiable criterion function, as previously discussed in \cite{beran_08}.




\section{Acknowledgments} \label{sec: ack}
The initial motivation for this research grew out of discussion with Lutz D\"umbgen while preparing exercises for
his lecture ``Optimization'' during summer semester 2006 at the University of Bern.
I thank Leonhard Held for discussions about the Bayesian perspective of the problem, Sarah Haile for proofreading the
final version, and my former employer,
the Swiss Group for Clinical Cancer Research (SAKK), for the permission to use the data of \cite{taussky_05}.

\texttt{R}-functions \citep{R} to efficiently compute $\ve{\hat \beta}$ for linear, logistic, and Cox-Regression
are bundled in the \proglang{R} package \pkg{OrdFacReg} \citep{OrdFacReg} and available from CRAN.

\appendix
\section{Details of the active set algorithm} \label{sec: details algo}
In this section, we complement the description of the algorithm indicated in Section~\ref{sec: general}.
Recall the sets of indices $\JJcp=\{c+1,\ldots,p\}$ and $\LLj = \{2, \ldots, k_j \}$ for $j \in \JJcp$. 

In order to respect the ordinal character of each of the factors $\mat{w}_{\cdot j}$ we introduced
in Section~\ref{sec: general} the new data matrix $\mat{X}$ by adding dummy variables for the ordered factors, such that
\bea
    \mat{X} &=& \Bl(\mat{w}_{\cdot 1}, \ldots, \mat{w}_{\cdot c} \, , \, \ve{x}_{\cdot \psi(j, l)} \Br)_{l \in \LLj; \ j \in \JJcp}
\eea for dummy variables \bea
    \ve{x}_{\cdot \psi(j, l)} 
    &=& (1 \{w_{ij} = l\})_{i=1}^n, \ l \in \LLj, \ j \in \JJcp.
\eea The function $\psi$ is given in \eqref{eq: def psi}.
With the above version of coding, $l=1$ is considered the reference level for every ordered factor
$\mat{w}_{\cdot j}$ and the resulting design matrix $\mat{X}$ is now an element of $\R^{n \times d}$ where
\bea
   d &=& \sum_{j \in \JJcp} \sum_{l \in \LLj} 1 \nonumber\\
     &=& c + \psi(p, k_p) - \psi(c+1, 2) +1 \nonumber\\
     &=& c-f + \sum_{j \in \JJcp} k_j. \label{eq: d}
\eea Again, we denote by $\ve{x}_{i \cdot}$ the $i$-th row of $\mat{X}$, \ie the values of the ``dummyfied''
predictors for the $i$-th observation.
In order to respect the ordinal character of each of the factors $\mat{w}_{\cdot j}$ we then constrain
optimization of the updated functional $L = L(\ve{y}, \mat{X}, \ve{\beta})$ to the space of
parameters $\BB(c, \, p, \, \ve{k})$ given in \eqref{eq: general B}.

We write $\ell$ as placeholder for any of the functions $\ell_{n,1}, \ell_{n,2},$ or $\ell_{n,3}$
(for ease of notation we omit the dependence on $n$) and the aim is to find for given response vector and matrix
of predictors the vector
\bea
    \ve{\hat \beta} &:=& \argmax_{\ve{\beta} \, \in \, \BB} \ell(\ve{\beta}).
\eea To fit the constrained maximization problem (\ref{eq: general constr opt problem}) into the framework of
\cite{duembgen_07}, we write the set $\BB$ given in (\ref{eq: general B}) as
\bea
    \BB &=& \{\ve{\beta} \in \R^d \ :  \ve{v}_i^{\trans} \, \ve{\beta}\le 0, \ i = 1, \ldots, q\}
\eea
for vectors $\ve{v}_i\in \R^d$. For ease of notation, we have enumerated the constraining inequalities
\renewcommand{\arraystretch}{1.6}
\bea
  \begin{array}{lclcl}
    \ve{v}_1^{\trans} \ve{\beta}         & = & - \beta_{c+1, \, 2}                           & \le & 0       \\
    \ve{v}_2^{\trans} \ve{\beta}         & = & - \beta_{c+1, \, 3} + \beta_{c+1, \, 2}       & \le & 0       \\
                                         &\vdots&                                            &\vdots&          \\
    \ve{v}_{k_{c+1}-1}^{\trans} \ve{\beta}   & = & - \beta_{c+1, \, k_{c+1}} + \beta_{c+1, \, k_{c+1}-1} & \le & 0 \\
    \ve{v}_{k_{c+1}}^{\trans} \ve{\beta}     & = & - \beta_{c+2, \, 2}                           & \le & 0   \\
    \ve{v}_{k_{c+1} + 1}^{\trans} \ve{\beta} & = & - \beta_{c+2, \, 3} + \beta_{c+2, \, 2}       & \le & 0 \\
                                         &\vdots&                                            &\vdots&          \\
    \ve{v}_q^{\trans} \ve{\beta}         & = & - \beta_{p, \, k_p} + \beta_{p, \, k_{p-1}}   & \le & 0
  \end{array}
\eea
\renewcommand{\arraystretch}{1}
from $i=1, \ldots, q$, where \bea
    q&=&\Bl(\sum_{j \in \JJcp} k_j\Br) - f.
\eea The function $\phi: \bigl\{(c+1) \times \LL_{c+1},\ldots,p \times \LL_p \bigr\}\to \{1, \ldots, q\}$ that
maps the original indices $(j, \, l)$
to the ``inequality index'' $i$ is given by
\bean
    \phi(j, \, l) &=& \Bl(\sum_{h=c+1}^{j} k_{h-1}\Br)+(l-1)-(j-c-1) \nonumber \\
    &=& \psi(j,l)-c \label{eq: def phi}
\eean so that the inequalities can be written as \bea
    \ve{v}_{\phi(j, \, l)}^{\trans}\ve{\beta} &=& - \beta_{j, \, l} + \beta_{j, \, l-1}1_{\{l \ge 3\}} \\
    &\le& 0
\eea for $l \in \LL_p$ and $j \in \JJcp$. The vectors $\ve{v}_i$ for any $i = \phi(j, \, l) \in \{1, \ldots,
q\}$ are received via \bea
   \ve{v}_i &:=& \Bl( 1\{k = c + \phi(j, l) - 1\}1\{l \ge 3\} - 1\{k = c + \phi(j, l)\} \Br)_{k=1}^q.
\eea Note that all these vectors are linearly independent. Define for any index set $A \subseteq \{1, \ldots,
q\}$ the linear subspace
\bea
    \VV(A) &:=& \Bl\{\ve{\beta} \in \R^d \ : \ \ve{v}_a^{\trans} \ve{\beta} = 0, \ \text{ for all } a \in A\Br\} \\
    &=& \Bl\{\ve{\beta} \in \R^d \ : \ -\beta_{j, \, l} + \beta_{j, \, l - 1} 1\{l \ge 3\} = 0, \ \text{ for all } j, \, l \text{ such that } \phi(j, \, l)\in A\Br\}
\eea and for $\ve{\beta} \in \R^d$ the set $A$ of ``active constraints'': \bea
    A(\ve{\beta}) &:=& \Bl\{i \in \{1,\ldots,q\} \ : \ \ve{v}_i^{\trans} \ve{\beta} \ge 0\Br\}.
\eea

\paragraph{Maximization on subspace}The crucial assumption for an active set algorithm is that we have an algorithm
available that for any $A \subseteq \{1, \ldots, q\}$ (efficiently) computes
\bea
    \ve{ \til\beta}(A) &=& \argmax_{\ve{\beta} \, \in \, \VV(A)} \ell(\ve{\beta}), \label{eq: max on subspace2}
\eea provided that $\VV(A) \cap \{\ve{\beta} \ : \ \ell(\ve{\beta}) > -\infty\} \ne 0$, see Section~\ref{sec: active set}.
For simplicity and without loss of generality, fix $j = c+1$.
Then, for a given $\ve{\beta}$ the following situations can cause a non-empty set $\VV(A)$:

\renewcommand{\arraystretch}{1.6}
\begin{table}[h]
{\footnotesize
\centerline{
\begin{tabular}{clll}
  Case & Violation(s) \hspace{2cm}                                     & $A(\ve{\beta})$ & Corresponding set $\VV(A)$ \\ \hline
  1    & $\beta_{c+1, \, 3}>\beta_{c+1, \, 2}, \, \beta_{c+1, \, 2}<0$ & $\{1\}$         & $\{\ve{\beta} \in \R^d \ : \ \ve{v}_1^{\trans} \ve{\beta} = 0\}$ \\
  2    & $\beta_{c+1, \, 2}>\beta_{c+1, \, 3}, \, \beta_{c+1, \, 2}>0$ & $\{2\}$         & $\{\ve{\beta} \in \R^d \ : \ \ve{v}_2^{\trans} \ve{\beta} = 0\}$ \\
  3    & $\beta_{c+1, \, 2}>\beta_{c+1, \, 3}, \, \beta_{c+1, \, 2}<0$ & $\{1, \, 2\}$   & $\{\ve{\beta} \in \R^d \ : \ \ve{v}_1^{\trans} \ve{\beta} = 0, \, \ve{v}_2^{\trans} \ve{\beta} = 0\}$ \\
  \end{tabular}
}
\vspace*{-0.15cm}
\caption{Possible violations of constraints within one ordered factor.} \label{table: possible violations}
}
\end{table}
\renewcommand{\arraystretch}{1}

Note that the situation $\ve{v}_s\ve{\beta}^{\trans} > 0$ for any $s = 3,\ldots, k_{c+1}$ can be treated analogously to
Case 2 in Table~\ref{table: possible violations}.

To compute the unrestricted maximizer $\ve{\til\beta}(A)$ in the three cases given in
Table~\ref{table: possible violations}, the strategy is to suitably modify the design matrix $\mat{X}$. Precisely,
we show how to construct new data matrices $\mat{X}_*^i$ and a
new corresponding function $\ell_*^i, i = 1, 2, 3: \R^{d_*} \to \R$ (here, $i$ stands for the corresponding
case in Table~\ref{table: possible violations}) for a given $A_* \subset \{1,\ldots,q\}$ in the three cases of
Table~\ref{table: possible violations} such that $\ve{ \til\beta}(A_*)$ can be immediately derived from
\bean
    \ve{\hat \beta_*}^i &=& \argmax_{\ve{\beta} \in \R^{d_*^i}} \ell_*^i(\ve{\beta}). \label{eq: opt problem on subspace}
\eean It is crucial to realize that the maximization in (\ref{eq: opt problem on subspace}) is unconstrained and the following
arguments show that $d_*^i \le d$ in all considered cases. In what follows, we explicitly state the unconstrained
maximization problem, assuming that only the case under consideration is present. Apparent combinations of these basic
strategies are necessary in case more than one of the three cases described in Table~\ref{table: possible violations}
are present.

\paragraph{Case 1} Writing down the maximization problem~(\ref{eq: opt problem on subspace}) explicitly, we get
\bea
     \ve{\til\beta}(\{1\}) &=& \argmax_{\beta_{c+1,\,2} = 0, \, \ve{\beta} \in \R^d}  \ell(\ve{\beta}) \\
                           &=& \Bl((\ve{\hat \beta_*}^1)_{i=1}^{c}, \, 0, \, (\ve{\hat \beta_*}^1)_{i=c+1}^{d-1}\Br)
\eea with
\bea
    \ve{\hat \beta_*}^1 &=& \argmax_{\ve{\beta} \in \R^{d-1}}\ell_*^1(\ve{\beta}, \mat{X}_{-(c+1)}),
\eea
where in general $\mat{M}_{-i}$ is the matrix $\mat{M}$ with the $i$-th column omitted and $\ell_*^1(\cdot, \mat{Q})$ is
the criterion function corresponding to $\ell$, but based on the design matrix $\mat{Q}$.

\paragraph{Case 2} Roughly, the strategy here is to add up the dummy variables corresponding to the violating
constraints, compute the unconstrained maximizer and then ``blow up'' the resulting estimator again. To see this,
consider
\bea
     \ve{\til\beta}(\{2\}) &=& \argmax_{\beta_{c+1,\,3} = \beta_{c+1,\,2}, \, \ve{\beta} \in \R^d}  \ell(\ve{\beta}) \\
                           &=& \Bl((\ve{\hat \beta_*}^2)_{i=1}^{c}, \, \hat \beta^2_{* \ c+1}, \, \hat \beta^2_{* \ c+1}, \, (\ve{\hat \beta_*}^2)_{i=c+2}^{d-1}\Br)
\eea with
\bea
    \ve{\hat \beta_*}^2 &=& \argmax_{\ve{\beta} \in \R^{d-1}}\ell_*^2\Bl(\ve{\beta}, \Bl((\mat{X})_{\cdot, \, i=1}^c, \mat{X}_{\cdot, \, (c+1)} + \mat{X}_{\cdot, \, (c+2)}, (\mat{X})_{\cdot, \, i=c+3}^d\Br) \Br).
\eea

\paragraph{Case 3} Repeating the above computations, we derive
\bea
     \ve{\til\beta}(\{1, \, 2\}) &=& \argmax_{\beta_{c+1,\,3} = \beta_{c+1,\,2} = 0, \, \ve{\beta} \in \R^d}  \ell(\ve{\beta}) \\
                                 &=& \Bl((\ve{\hat \beta_*}^3)_{i=1}^{c}, \, 0, 0, \, (\ve{\hat \beta_*}^3)_{i=c+1}^{d-2}\Br)
\eea where
\bea
    \ve{\hat \beta_*}^3 &=& \argmax_{\ve{\beta} \in \R^{d-2}}\ell_*^3(\ve{\beta}, \mat{X}_{-(c+1, \, c+2)}).
\eea


\section{Proofs} \label{sec: proofs}

\paragraph{Proof of Lemma~\ref{lem: simple case PAVA}} First, observe that for $i = 1,\ldots, n$,
\bea
    1\{w_{i1} = q \} 1\{w_{i1} = r\} &=& 0 \ \text{ for } \ 2 \le q, r \le k_1 \ \text{ with } \ q \ne r.
\eea The function $-\ell_{n,1}$ can then be written as
\bea
    -\ell_{n,1} (\ve{\beta}) 
    &=& \sumn \Bl( y_i - \sum_{l=2}^{k_1} \beta_l 1\{x_{il} = 1\} \Br)^2 \\
    &=& \sumn \Bl(y_i^2-2y_i \sum_{l=2}^{k_1} \beta_l 1\{x_{il} = 1\} + \Bl(\sum_{l=2}^{k_1} \beta_l 1\{x_{il} = 1\}\Br)^2\Br) \\
    &=& \sumn y_i^2 - 2 \sum_{l=2}^{k_1} \beta_l \sumn y_i 1\{x_{il} = 1\} + \sum_{l=2}^{k_1} \beta_l^2 \sumn 1\{x_{il} = 1\} \\
    &=& \sum_{l=2}^{k_1} \Bl( \beta_l^2 N_l -2\beta_l \sum_{i : x_{il} = 1} y_i\Br) + \sumn y_i^2 \\
    &=& \sum_{l=2}^{k_1} N_l \Bl(\beta_l^2 - 2 \beta_l \sum_{i : x_{il} = 1} y_i / N_l \Br) + \sumn y_i^2 \\
    &=& \sum_{l=2}^{k_1} N_l \Bl((\beta_l - m_l )^2 - m_l^2\Br)+\sumn y_i^2 \\
    &=& \sum_{l=2}^{k_1} N_l (\beta_l - m_l )^2 + \mathrm{const}(\ve{y}, \mat{X}).
\eea The minimum of the latter expression under the constraint $\beta_2 \le \ldots \le \beta_{k_1}$ can
easily be found using PAVA. \hfill $\Box$

\paragraph{Proof of Theorem~\ref{lem: characterization}}
Before coming to the actual proof, we state a necessary lemma.
\begin{lemma} \label{lem: general a b}
Let $\ve{a}, \ve{b} \in \R^n$ be two vectors having the following properties: \bean
    \sum_{i=1}^j a_i &\ge& 0_{\color{white}i} \ \text{ for all } \ j = 1, \ldots, n \label{eq: ab1} \\
    \sum_{i=k}^n a_i &\le& 0_{\color{white}i} \ \text{ for all } \ k = 1, \ldots, n \label{eq: ab2} \\
    b_{i} &\ge& b_{i-1}  \ \text{ for all } \ i = 2, \ldots, n                      \label{eq: ab3} .
\eean Then \bea
    \sum_{i=1}^n a_i b_i &\le& 0.
\eea
\end{lemma}

First, we prove that if $\ve{\hat \gamma}_1$ maximizes $\ell$ over $\BB$, then (\ref{eq: charac1})-(\ref{eq:
charac2}) are fulfilled. To this end, let $t > 0$ be small enough and let $\ve{\Delta} \in \R^d$ be a vector such that
$\ve{\hat \gamma}_1 + t \ve{\Delta} \in  \BB$. Since $\ve{\hat \gamma}_1$ maximizes the concave function $\ell$ we have
\bea
    \frac{\d}{\d t} \ell(\ve{\hat \gamma}_1+t\ve{\Delta})\vert_{t=0} &\le& 0,
\eea which entails \bean
    \nabla \ell(\ve{\hat \gamma}_1)^{\trans} \ve{\Delta} &\le&0. \label{eq: charac Delta}
\eean We then get (\ref{eq: charac1})-(\ref{eq: charac3}) using the following perturbation functions:
\bea
    \ve{\Delta}_1 = \ve{\Delta}_1(c)          &=& \pm  (1\{s \le c \})_{s=1}^d, \\
    \ve{\Delta}_2 = \ve{\Delta}_2(j, \ve{h}_j, t)  &=& - \Bl(1\{s = \underline{B_3}, \ldots, t\}\Br)_{s=1}^d  \\ 
    \ve{\Delta}_3 = \ve{\Delta}_3(j, \ve{h}_j, t)  &=& \Bl(1\{s = t, \ldots, \overline{B_3}\}\Br)_{s=1}^d
\eea
for all $t \in B_{3, j, u}(\ve{\hat \gam}), \, j \in \JJcp, \text{ and } u = 1,\ldots, |\ve{h}_j|$
and where we defined $\overline{B_3} = \max B_{3, j, u}(\ve{\hat \gam})$ and
$\underline{B_3} = \min B_{3, j, u}(\ve{\hat \gam})$.
Now suppose we are given a vector $\ve{\hat \gamma}_2$ that fulfills (\ref{eq: charac1})-(\ref{eq: charac3}).
We then have to show that \bea
    \ve{\hat \gamma}_2 &=& \argmax_{\ve \beta \in \BB} \ell(\ve{\beta}).
\eea From convex analysis, it is well known that this is equivalent to show
\bean
    \lim_{t \searrow 0} \frac{\ell(\ve{\hat \gamma}_2 + t(\ve{g} - \ve{\hat \gamma}_2)) - \ell(\ve{\hat \gamma}_2)}{t} &=& \lim_{t \searrow 0} \frac{\ell(\ve{\hat \gamma}_2 + t\ve{\Delta}) - \ell(\ve{\hat \gamma}_2)}{t} \nonumber\\
    &=& \nabla \ell(\ve{\hat \gamma}_2)^{\trans} \ve{\Delta} \\
    &\le&0 \label{eq: to show charact} \label{eq: to show charact2}
\eean for arbitrary vectors $\ve{\Delta} = \ve{g}-\ve{\hat \gamma}_2$ such that $\ve{g} \in \BB$. Now compute
\bean
    \nabla \ell(\ve{\hat \gamma}_2)^{\trans} \ve{\Delta} & = & \sum_{i \in }\nabla \ell(\ve{\hat \gamma}_2)^{\trans} (\ve{g}-\ve{\hat \gamma}_2) \nonumber \\
    &=& \sum_{j \in \JJcp} \sum_{u=1}^{|\ve{h}_j|} \sum_{s \in B_{3, j, u}} \Bl( g_s (\nabla \ell(\ve{\hat \gamma}_2))_s - (\ve{\hat \gamma}_2)_s (\nabla \ell(\ve{\hat \gamma}_2))_s\Br) \nonumber \\
    &=& \sum_{j \in \JJcp} \sum_{u=1}^{|\ve{h}_j|} \sum_{s \in B_{3, j, u}} g_s (\nabla \ell(\ve{\hat \gamma}_2))_s  - \sum_{j \in \JJcp} \sum_{u=1}^{|\ve{h}_j|} (\ve{\hat \gamma}_2)_s \sum_{s \in B_{3, j, u}} (\nabla \ell(\ve{\hat \gamma}_2))_s \label{eq: theo toshow}.
\eean The second term disappears due to \eqref{eq: charac4}. As for the first term, we invoke
Lemma~\ref{lem: general a b} where $\nabla \ell(\ve{\hat \gamma}_2)$ takes the role of $\ve{a}$ and $\ve{g}$ that of
$\ve{b}$ to finally deduce that \eqref{eq: theo toshow} is at most 0. \hfill $\Box$

\paragraph{Proof of Lemma~\ref{lem: general a b}}
First, note that (\ref{eq: ab1}) and (\ref{eq: ab2}) immediately imply
\bea
    \sum_{i=1}^n a_i & = & 0. \label{eq: ab4}
\eea
Using this, one deduces
\bea
    \sum_{i=1}^n a_i b_i & = & \sum_{i=2}^n a_i (b_i-b_1) \\
                         & = & \Bl(\sum_{i=2}^{n-1} a_i (b_i-b_1)\Br) + a_n (b_n-b_1) \\
                         &\le& \Bl(\sum_{i=2}^{n-1} a_i (b_i-b_1)\Br) + a_n (b_{n-1}-b_1) \ \text{since } \ a_n \le 0 \ \text{ and due to \eqref{eq: ab3}} \\
                         & = & \Bl(\sum_{i=2}^{n-2} a_i (b_i-b_1)\Br) + (a_{n-1}+a_n) (b_{n-1}-b_1) \\
                         &\le& \Bl(\sum_{i=2}^{n-2} a_i (b_i-b_1)\Br) + (a_{n-1}+a_n) (b_{n-2}-b_1) \ \text{ due to \eqref{eq: ab2} and \eqref{eq: ab3}} \\
                         &\le& \Bl(\sum_{i=2}^{n-3} a_i (b_i-b_1)\Br) + (a_{n-2}+a_{n-1}+a_n) (b_{n-2}-b_1).
\eea Repeatedly applying this same trick we finally arrive at
\bea
    \sum_{i=1}^n a_i b_i & = & a_2(b_2-b_1) + \Bl(\sum_{i=3}^n a_i\Br) (b_3-b_1) \\
                         &\le& \Bl(\sum_{i=2}^n a_i\Br) (b_2-b_1).
\eea By means of \eqref{eq: ab2} and \eqref{eq: ab3} the latter expression remains non-positive. \hfill $\Box$


\bibliographystyle{ims}
\bibliography{ordreg}

\end{document}